\documentclass[aps,showpacs,prl]{revtex4}
\usepackage{epsfig,amssymb,amscd,graphicx}
\usepackage{subfigure}

\def\bra#1{\mathinner{\langle{#1}|}}
\def\ket#1{\mathinner{|{#1}\rangle}}

\begin{document}

\title{Dynamical Properties of the Delta Kicked Harmonic Oscillator}

\author{G. A. Kells}\email{gkells@thphys.may.ie}
\author{J. Twamley}
\author{D. M. Heffernan} 
\affiliation{Dept. of Mathematical Physics, National University of Ireland, Maynooth,  Ireland.}

\begin{abstract}
We propose an efficient procedure for numerically evolving the quantum dynamics of Delta Kicked Harmonic Oscillator. The method allows for longer and more accurate simulations of the system as well as a simple procedure for calculating the system's Floquet eigenstates and quasi-energies. The method is used to examine the dynamical behaviour of the system in cases where the ratio of the kicking frequency to the system's natural frequency is both rational and irrational. 
\end{abstract}

\pacs{05.45.Pq, 03.65.-w, 05.45.Mt}

\date{\today}
\maketitle

Kicked Hamiltonian systems are firmly established as prototypical models for studying chaos. Enormous effort has been spent examining models like the standard map and the kicked rotator. The study of kicked degenerate systems, those that do not obey the KAM theorem, have received relatively little attention but display fascinating dynamical properties both classically and quantum mechanically. The Kicked Harmonic Oscillator (KHO) is an example of such a system. The system was originally proposed as a 2-dimensional model of kicked charges in a uniform magnetic field \cite{Zaslavsky}. It has also been proposed as a model for electronic transport in semiconductor superlattices \cite{Fromhold} and atom optic modeling using ion-traps \cite{Gardiner}. In particular, the quantum kicked harmonic oscillator, in the same way as the quantum kicked rotator \cite{Levi}, may be simulated efficiently on a quantum information processor using the Quantum Fractional Fourier Transform \cite{Klappenecker}.

The Hamiltonian of the system may be written as

\begin{equation}
H_{KHO}=\omega_{HO} \left( \frac{ p^{2}}{2}+ \frac{ q^{2}}{2}\right)+V(q)\sum_{n=-\infty}^{\infty}\delta(t-n T_{K})
\label{HKHO}
\end{equation}

\noindent
where the kicking potential $V(q)$ in our system is given by

\begin{equation}
V(q)= \mu \cos(kq)
\label{V}
\end{equation}

\noindent
where k is a parameter related to $\omega_{HO}$ and the mass $m$. It is distinct from other widely studied systems because of the presence of two different frequencies in the Hamiltonian, namely the natural frequency of the Harmonic Oscillator $\omega_{HO}$, and the kicking frequency $\omega_K=2 \pi/T_K$. It is the ratio between these frequencies, $\omega_{HO}/\omega_K = 1/R$, and the strength of the kicking potential, that determines the global behavior of the system \cite{Zaslavsky}. It is appropriate to note here that the kicking potential we use is even. Using an odd kicking potential, say $V(q)= \mu \sin(kq)$, changes the global behaviour of the system dramatically \cite{Dana2}. We do not attempt to analyse this case although the numerical techniques that we present later can be applied to any delta kicking potential.

Understanding the quantum system analytically has proved to be problematic. While expressions have been found for the Floquet operator \cite{Daly}, these generally involve complicated infinite summations of Bessel functions and are too cumbersome to yield any important information about the dynamics of the system. Numerical simulations are therefore essential to understanding how the system evolves. However, in general, factors such as memory and computation time reduce the accuracy of simulations before much can be concluded about the eigensolutions or long time dynamical evolution of the system. We outline a procedure, based on the Fractional Fourier Transform, that allows for efficient numerical simulation of the system's evolution as well as an effective and straightforward approach to calculating the system's Floquet operator. We use this approach to confirm previous results for the system where $R=4,5$ as well as showing that delocalisation can occur when $R$ is irrational if the kicking potential is strong enough.

The Floquet operator for this system can be factorized into two parts: 

\begin{equation}
U_{KHO}(T_K)=U_{SHO}(T_K)U_{kick}
\label{UKHO}
\end{equation} 

\noindent
where because of the delta function in Eq.(\ref{HKHO}) we can express the kick operator as

\begin{equation}
U_{kick}=\exp \left(\frac{-i }{\hbar}\mu \cos(kq)\right),
\label{Ukick}
\end{equation}

\noindent
which is clearly diagonal in the position basis and between kicks

\begin{equation}
U_{SHO}(T_K)=\exp \left( \frac{-i\omega_{HO} T_K (p^2 +q^2)}{2\hbar}\right),
\label{USHO}
\end{equation}

\noindent
which is diagonal in the Fock state basis but not in the position basis or the momentum basis. Changing from the number basis to the position basis requires a matrix-vector multiplication ($N^2$ operations) where $N$ is the number of dimensions being used to approximate the infinite dimensional Hilbert space. However, it is well known that changing between the position and momentum basis can be done through Fourier Transforms. This fact is used to great advantage in the split step method \cite{Feit}. In this method the exact operator $U_{SHO}$,applied for the small duration $\Delta t$, can be approximated by  

\begin{equation}
U_{SHO}'(\Delta t) = \exp \left(\frac{-i\omega_{HO} \Delta t p^2}{4 \hbar}\right)\exp \left(\frac{-i\omega_{HO}\Delta t q^2}{2\hbar}\right) \exp \left(\frac{-i \omega_{HO} \Delta t p^2}{4\hbar}\right)
\label{USHO'}
\end{equation}

\noindent
where we use the Fast Fourier Transform (FFT) \cite{Cooley}, to perform the change of basis. Although this method is not exact, the error is $O(\Delta t)^3$, and can therefore be made acceptably small by an appropriate choice of $\Delta t$. The number of operations needed to evolve the system from time $t$ to time $t+\Delta t$ is $O(N\log N)$. However, using the stability condition for the Split-Step method \cite{Weideman}, $ \Delta t < \frac{l^2}{\pi}$, where $l$ is the grid spacing, we roughly estimate the number of operations needed to accurately evolve the system over a finite time to be of $O(N^2 \log N)$. The essential point is that we avoid the use of matrix-vector multiplications through the use of Fast Fourier Transforms. This technique is relatively efficient computationally and does not require large amounts of storage.

It is possible however, to exploit certain symmetries in the Hamiltonian to evolve the system with far greater efficiency. The previous discussion regarded the Fourier Transform as a passive transformation, that is, one that changes the position basis to the momentum basis. However, for the harmonic oscillator, a Fourier Transform can also be regarded as an active transformation. In 1980, Namias defined the Fractional Fourier Transform in terms of the Hermite polynomials and showed that it is in fact the time evolution operator of the harmonic oscillator up to a phase \cite{Namias}. Explicitly, taking $\Psi(x,t)$ as the time-dependent  wavefunction of the harmonic oscilator, he showed that 
\begin{equation}
\Psi(x,t)=U_{SHO}(t)\Psi(x,0)=e^{-it/2} F_{-t} \Psi(x,0),
\label{Namias}
\end{equation}

\noindent
where the operator $F_{-t}$ is a generalisation of the ordinary Fourier transform \cite{Defn}. Indeed the latter is just the special case when $t = \pi/2$. This result is important for our system because of the existence of Fast Fractional Fourier Transforms (FFrFT's) \cite{Ozaktas}. There are a number of algorithms in the literature for performing the Fast Fractional Fourier Transform as efficiently as the regular FFT \cite{code}. This means that, in the position basis, the evolution of the wavefunction over a finite time may be written as

\begin{equation}
\Psi(x,T_K) \equiv U_{KHO}(T_K)\Psi(x,0)
=
e^{-iV(x)/\hbar}e^{-i T_K/2} F_{-T_K}  \Psi(x,0) ,
\label{UF}
\end{equation}

\noindent
and can be performed in about $O(N\log N)$ operations. This allows us to perform highly efficient numerical simulations of pure state quantum dynamics in Hilbert space dimensions of about $2^{17}$. This enables us to follow the diffusing quantum evolution for larger times and in greater detail. Simulations of such a size would be prohibitively costly in memory and processor time using other simulation techniques.

As an additional advantage, the Floquet matrix itself can be calculated. This is  a consequence of Eq. (\ref{UF}) and the fact that the Fractional Fourier Transform matrix can be generated without much additional overhead \cite{Ozaktas}. This allows for quick and straightforward calculation of the Floquet eigenstates and eigenvalues using the standard LAPACK routines.

\vspace{5mm}

One of the interesting properties of the system given in (\ref{HKHO}) and (\ref{V}), is the presence of a stochastic web that spans all of the classical phase space for certain values of the parameter $R$. Indeed the values of $R$ for which this crystalline structure appears can be seen to be related to the tessellation of the phase plane, where the values of $R = 3,4,6 $ represent filling (tiling) the plane with triangles, squares or hexagons respectively (we ignore the degenerate cases of  $R = 1,2 $). In the quantum mechanical problem these specific values of $R$ have, for all values of the kicking strength, been shown to give rise to extended quantum eigenstates and therefore delocalisation of the quantum state over time \cite{Berman,Borgonovi,Frasca}.

In the other cases where $R$ is an integer, a quasi crystalline structure is observed in the classical phase plane.In the quantum system there is a large suppression of the states quantum evolution with respect to energy \cite{Borgonovi,Shepelyansky}. These results do not appear to contradict those of the tight binding prediction \cite{Frasca}, where extended states are predicted for all rational values of the frequency ratio. Indeed, there is a value of the kick strength, above which, the quantum system diffuses at a rate, less than, but comparable to that of the classical system \cite{Shepelyansky}, whose energy is known to grow linearly with time.

In the case of irrational frequency ratios all traces of a crystalline structure disappear in the classical phase plane. The tight binding model predicts localisation for all values of the kick strength \cite{Frasca}, and this has been supported numerically for low kick strengths \cite{Borgonovi}. As far as we are aware there is no published numerical data for irrational frequency ratios for this system with large kick strength.

To study the localisation properties of the system we first investigate the mean energy after each application of the Floquet operator. We calculate 

\begin{equation}
E(t)=\bra{\psi_t}\frac{p^2}{2}+\frac{q^2}{2}\ket{\psi_t},
\end{equation}

\noindent
as a function of the discrete time $T_K$ or the kick number $n$. A direct comparison may be made between the quantum state and a classical ensemble. Defining $\alpha=(q+ip)/\sqrt{2\hbar}$ we have the wavefunction  of a coherent state centered at the coordinates $(q,p)$, 

\begin{equation}
\psi(q) = \left(\frac{\omega_0}{\pi \hbar}\right)^{1/4} \exp \left(-\frac{\omega_0}{2\hbar}q^2 +\sqrt{\frac{2\omega_0}{\hbar}}\alpha q-\frac{|\alpha|^2}{2}-\frac{\alpha^2}{2}\right).
\label{coherent}
\end{equation}

\noindent
The Wigner function of this state is a 2-D Gaussian that is centered about the point $(q,p)$ \cite{Hillery}. The Husimi function is very similar to this and can be regarded as the convolution of the Wigner function with a 2-D Gaussian centered at the origin. In the numerical simulations we compare the quantum evolution of a coherent state with the classical evolution of an ensemble of normally distributed particles initially centered around $(q,p)$ and possessing the same marginals as the corresponding Husimi function.

\begin{figure}[!ht]
\begin{center}
\setlength{\unitlength}{1cm}
\begin{picture}(15,15)
\put(-1,0){\includegraphics[width=15cm]{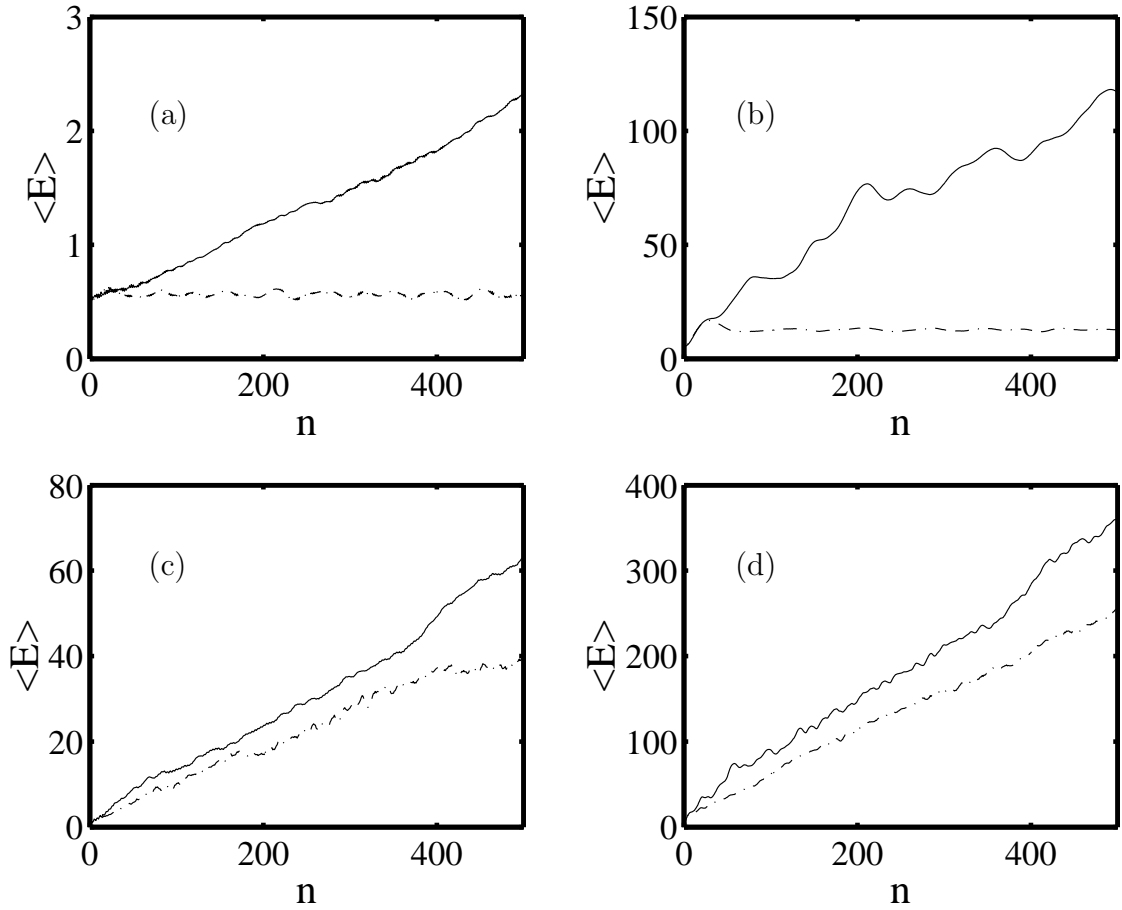}}
\put(1,10.5){\large(a)}
\put(8.8,10.5){\large(b)}
\put(1,4.5){\large(c)}
\put(8.8,4.5){\large(d)}
\end{picture}
\caption{\label{fig:1_4} Mean energy versus time for $\omega_{HO}/\omega_K =1/R= 1/4$ and $\hbar =1$. (a), (b) $\mu=0.5$ with initial state centered at $(0,0)$ and $(0,\pi)$ respectively. (c), (d) $\mu=2$ with initial state centered at $(0,0)$ and $(0,\pi)$ respectively. A Hilbert space of  dimension of $2^{16}$ was used in the simulation. Solid (Dashed) curves represent the quantum (classical) system.}
\end{center}
\end{figure}

When $R=4$, the quantum system can be shown analytically to be identical to the well studied Kicked Harper Model (KHM) \cite{Dana2}. In this case the quantum system diffuses for all values of the kick strength $\mu$, and for all initial conditions. To demonstrate this we center the initial state on a classical stable point $(q,p)=(0,0)$, and note that the energy of the quantum state grows with time (or kick number $n$), while the energy of the classical state remains bounded. From Fig. \ref{fig:1_4}(a), we see that the quantum system is able to tunnel through the separatrix which completely isolates classical trajectories. For completeness we also illustrate how both the quantum and classical systems diffuse when placed on an unstable point $(q,p)=(0,\pi)$, on the classical system's stochastic web. It is worth noting that at $\mu =0.5$, the stochastic web is too small to allow classical diffusion. However we see that the quantum system uses the separatrix to significantly increase its rate of diffusion. 

\begin{figure}[!ht]
\begin{center}
\setlength{\unitlength}{1cm}
\begin{picture}(12,12)
\put(-1,0){\includegraphics[width=15cm]{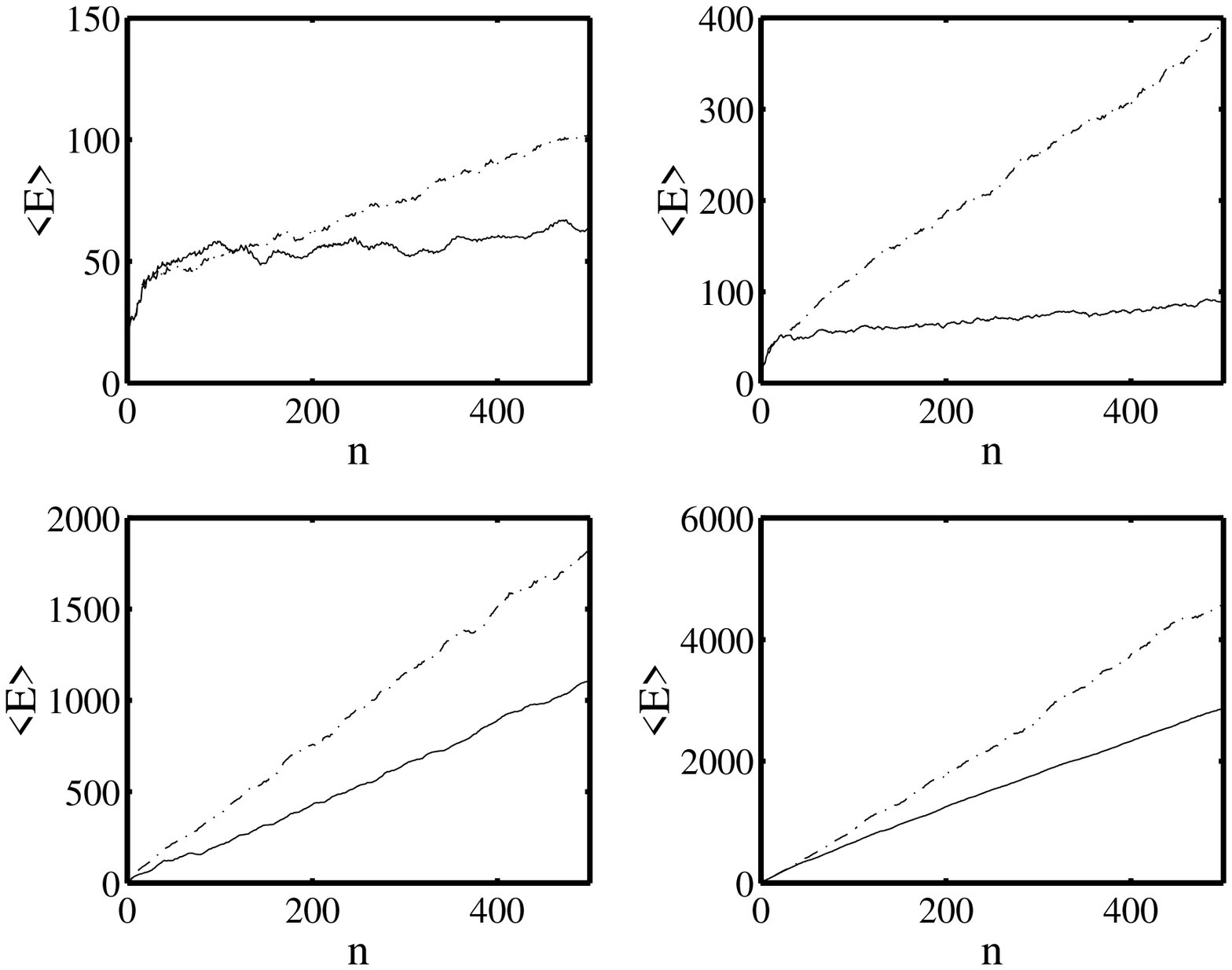}}
\put(1,10.5){\large(a)}
\put(8.8,10.5){\large(b)}
\put(1,4.5){\large(c)}
\put(8.8,4.5){\large(d)}
\end{picture}
\caption{\label{fig:1_5} Mean energy versus time for $\omega_{HO}/\omega_K =1/R= 1/5$ and $\hbar=1$. (a) $\mu=1$, (b) $\mu=2$, (c) $\mu=4$, (d) $\mu=6$. The initial state was centered at $(q,p)=(7.5,0)$. A Hilbert space of  dimension of  $2^{16}$ was used in this simulation. Solid (Dashed) curves represent the quantum (classical) system. } 
\end{center}
\end{figure}

In the case where $R=5$, the classical phase plane shows a quasi-crystalline structure. From  Fig. (\ref{fig:1_5}) we can see that at low kicking strengths diffusion of the quantum system is somewhat suppressed. However, above a certain critical value of the kick strength $\mu$, which depends on the initial position of the quantum state, this suppression no longer seems to exist.

\begin{figure}[!ht]
\begin{center}
\setlength{\unitlength}{1cm}
\begin{picture}(12,12)
\put(-1,0){\includegraphics[width=15cm]{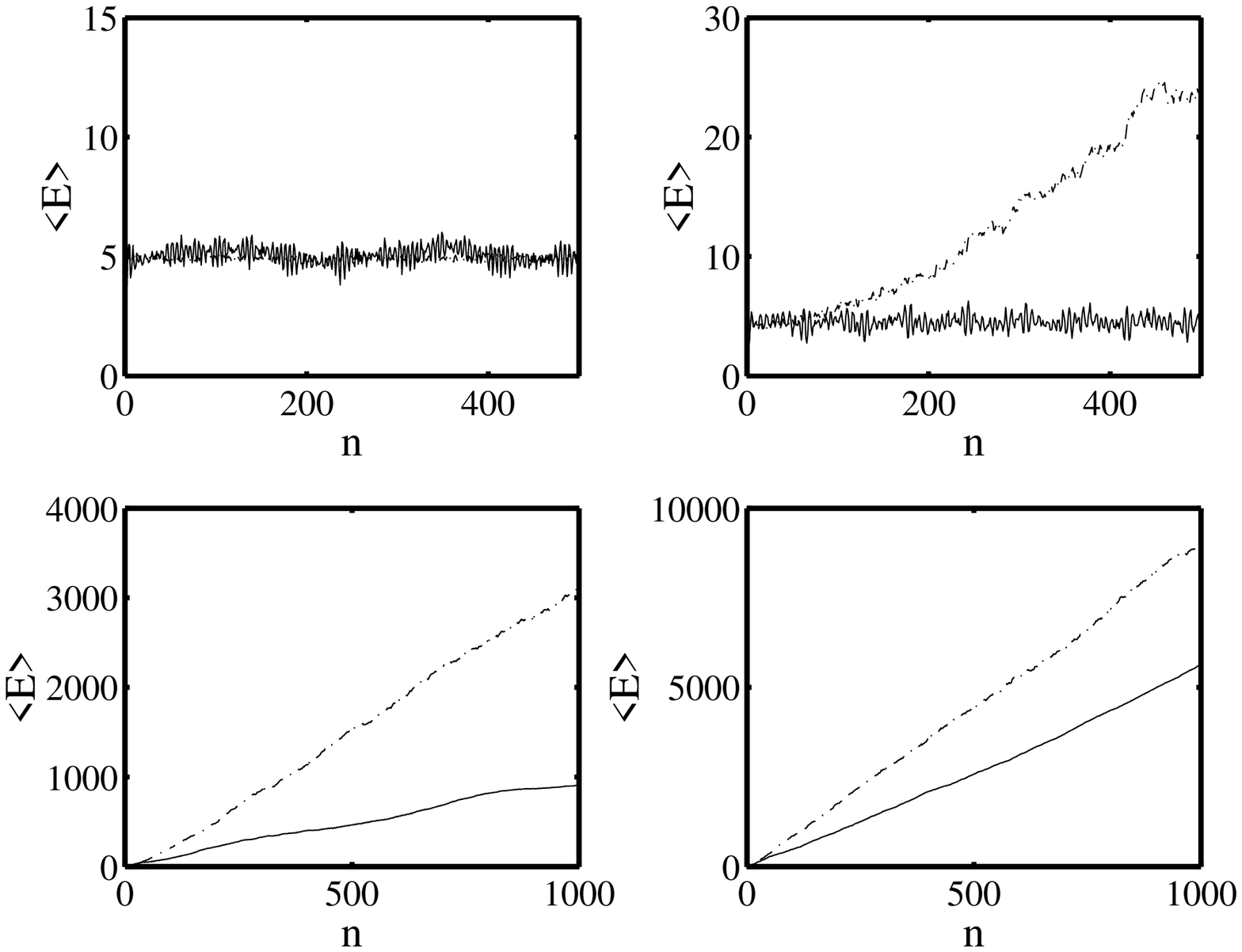}}
\put(1,10.5){\large(a)}
\put(8.5,10.5){\large(b)}
\put(1,4.5){\large(c)}
\put(8.5,4.5){\large(d)}
\end{picture}
\caption{\label{fig:irrational} Mean energy versus time for $\omega_{HO}/\omega_K = 1/R=(\sqrt{5}+1)/2$, and $\hbar=1$. (a) $\mu=1$, (b) $\mu=2$, (c) $\mu=4$, (d) $\mu=6$. The initial state was centered at $(q,p)=(\pi,0)$.A Hilbert space of  dimension of   $2^{17}$ was used for this simulation. Solid (Dashed) curves represent the quantum (classical) system.} 
\end{center}
\end{figure}

We see a similar type of global behavior when the frequency ratio is irrational even though the structure of the phase space patterns are quite different. Both the quantum and classical systems remain bounded up to different critical values of the kick strength $\mu$ (again the precise value of this parameter depends on the initial conditions). Above these values the the system begins to diffuse linearly. In the same way as before, there exists a range of kick strengths where the classical system diffuses and the quantum motion remain suppressed,see Fig.\ref{fig:irrational}(b)(c). It would appear that, in this region, the quantum state cannot flow along the same chaotic trajectories as the classical ensemble. This is opposite to the case for which $R=4$, where at low kick strengths, the quantum system appears to exploit the infinitesimal stochastic web to aid its dispersion, see Fig.\ref{fig:1_4}(a)(b).

\begin{figure}[!ht]
\begin{center}
\setlength{\unitlength}{1cm}
\begin{picture}(12,12)
\put(-0.5,5.5){\includegraphics[width=6cm]{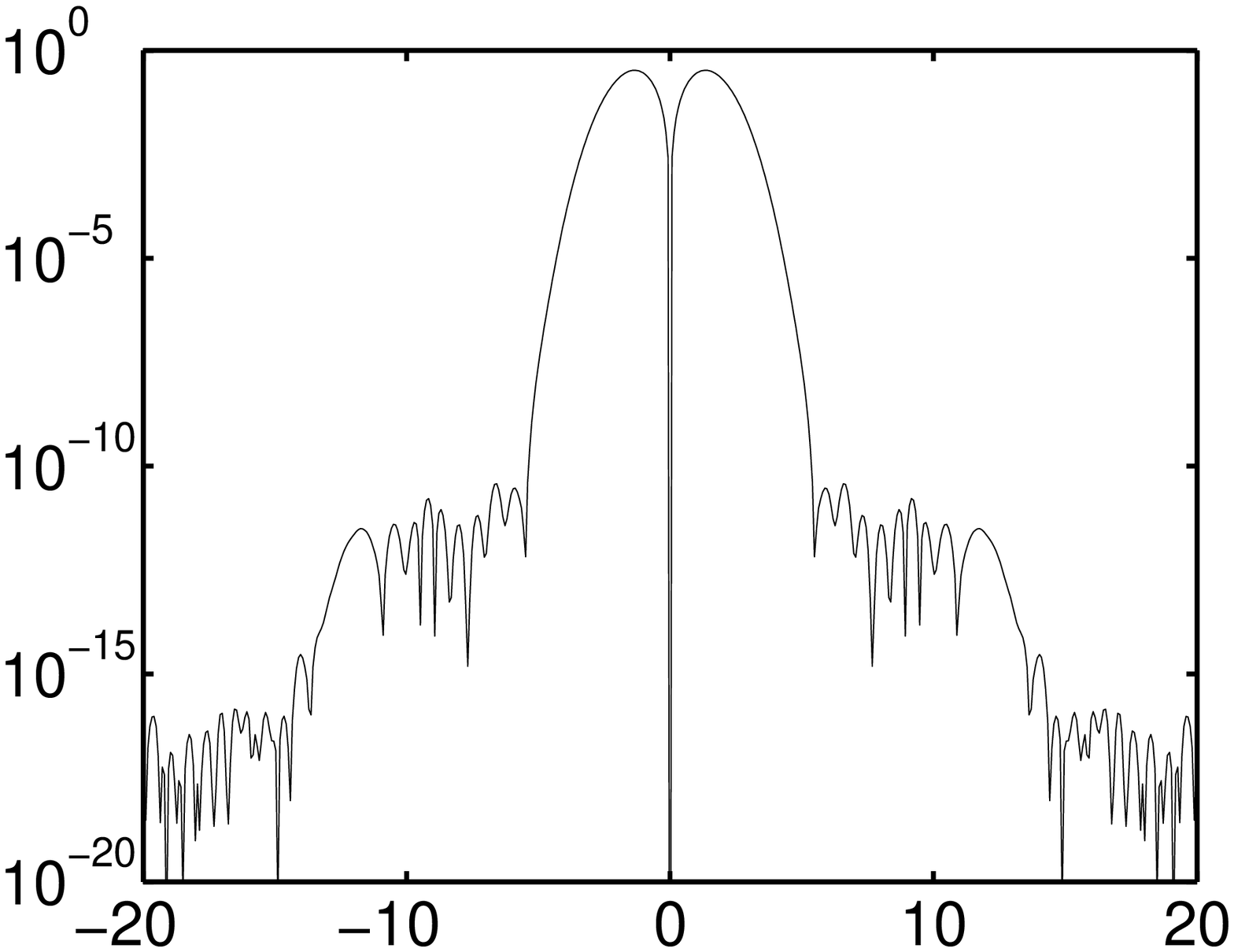}}
\put(-0.5,0.5){\includegraphics[width=6.5cm]{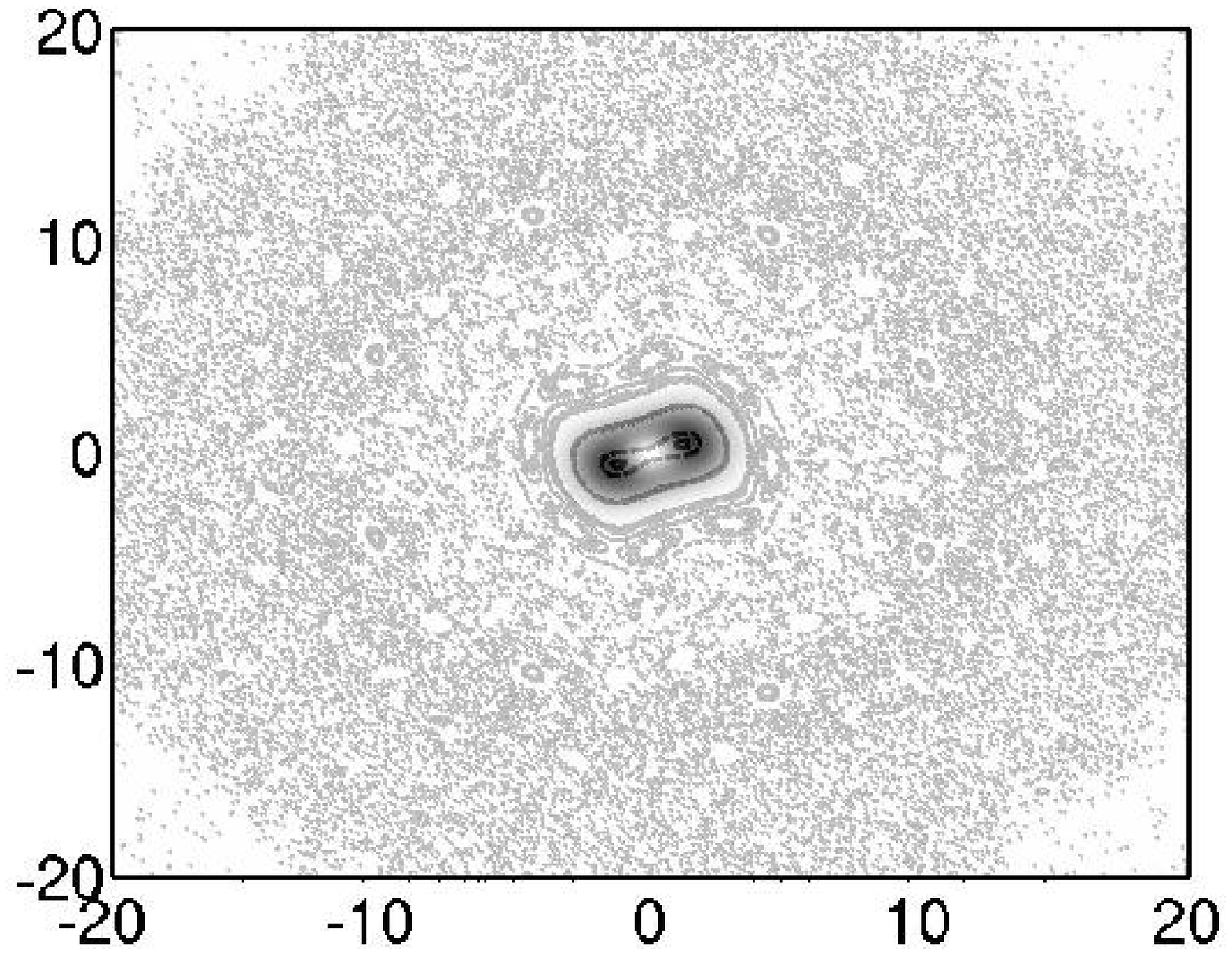}}
\put(6,5.5){\includegraphics[width=6cm]{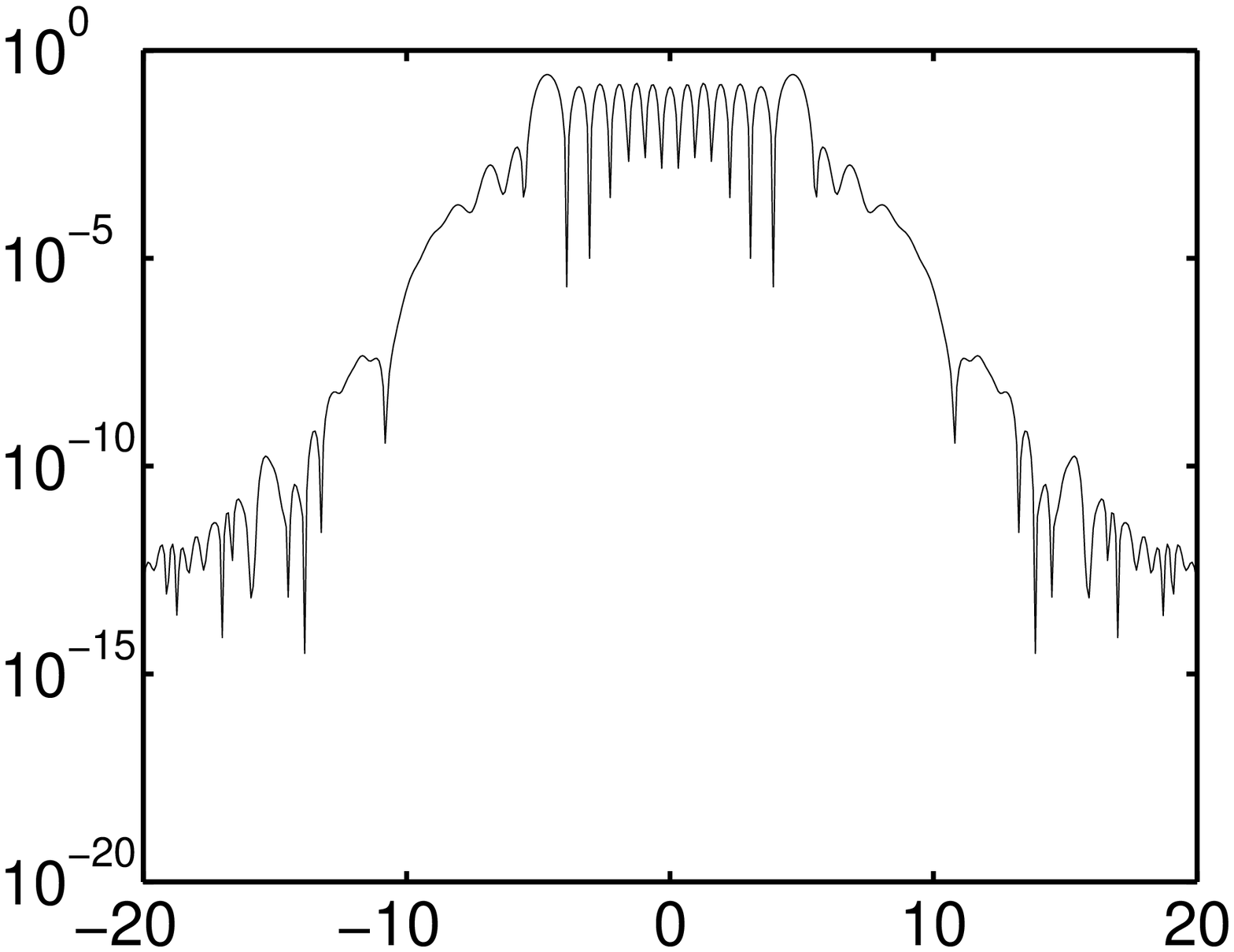}}
\put(6,0.5){\includegraphics[width=6.5cm]{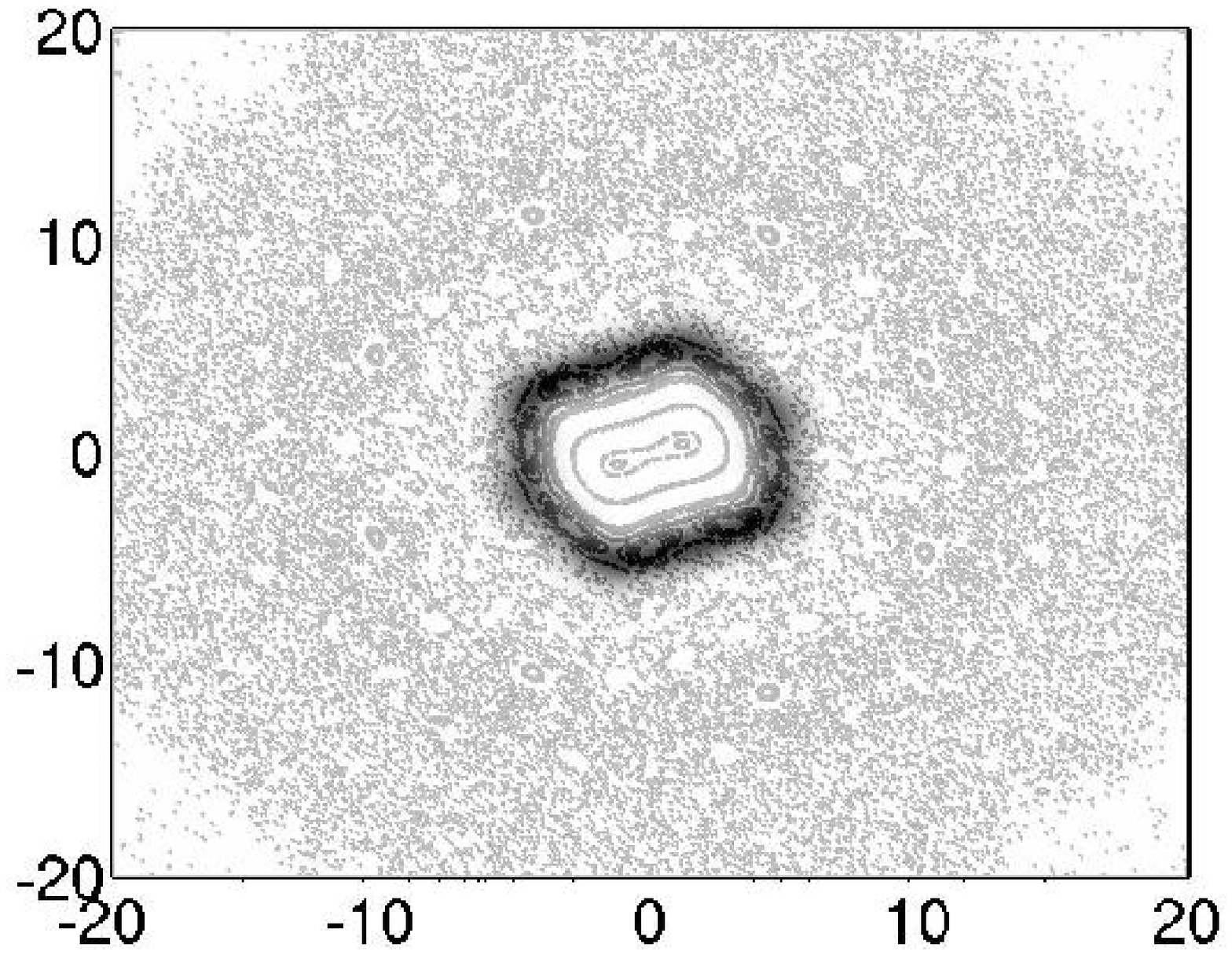}}
\put(0.5,9.5){\large(a)}
\put(7,9.5){\large(b)}
\put(0.5,4.5){\large(c)}
\put(7,4.5){\large(d)}
\put(-1.3,7.8){\large{$|\Psi|^2$}}
\put(-.7,3.0){\large{p}}
\put(2.75,0.4){\large{q}}
\put(9.25,0.4){\large{q}}
\end{picture}
\caption{\label{fig:states1} (a) and (b) Low energy eigenstates for the irrational system with kick strength $\mu = 1$, plotted on a logarithmic scale. Below each state, (c) and (d), are their Husimi functions plotted in overlay with the classical phase space Poincare plot. A Hilbert space of  dimension of  $2^{10}$ was used for these simulations. } 
\end{center}
\end{figure}

\begin{figure}[!ht]
\begin{center}
\setlength{\unitlength}{1cm}
\begin{picture}(12,12)
\put(-0.5,5.5){\includegraphics[width=6cm]{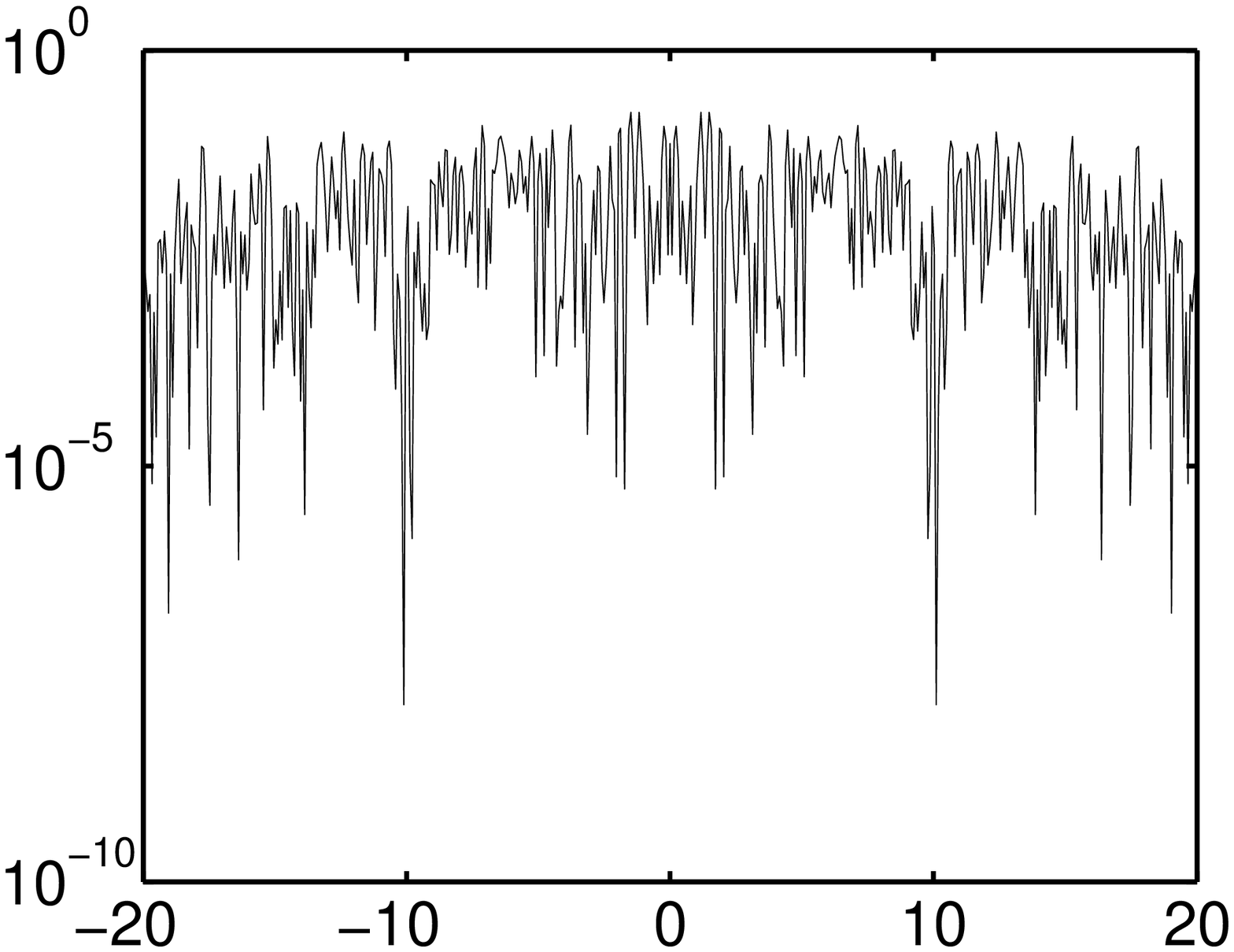}}
\put(-0.5,0.5){\includegraphics[width=6.5cm]{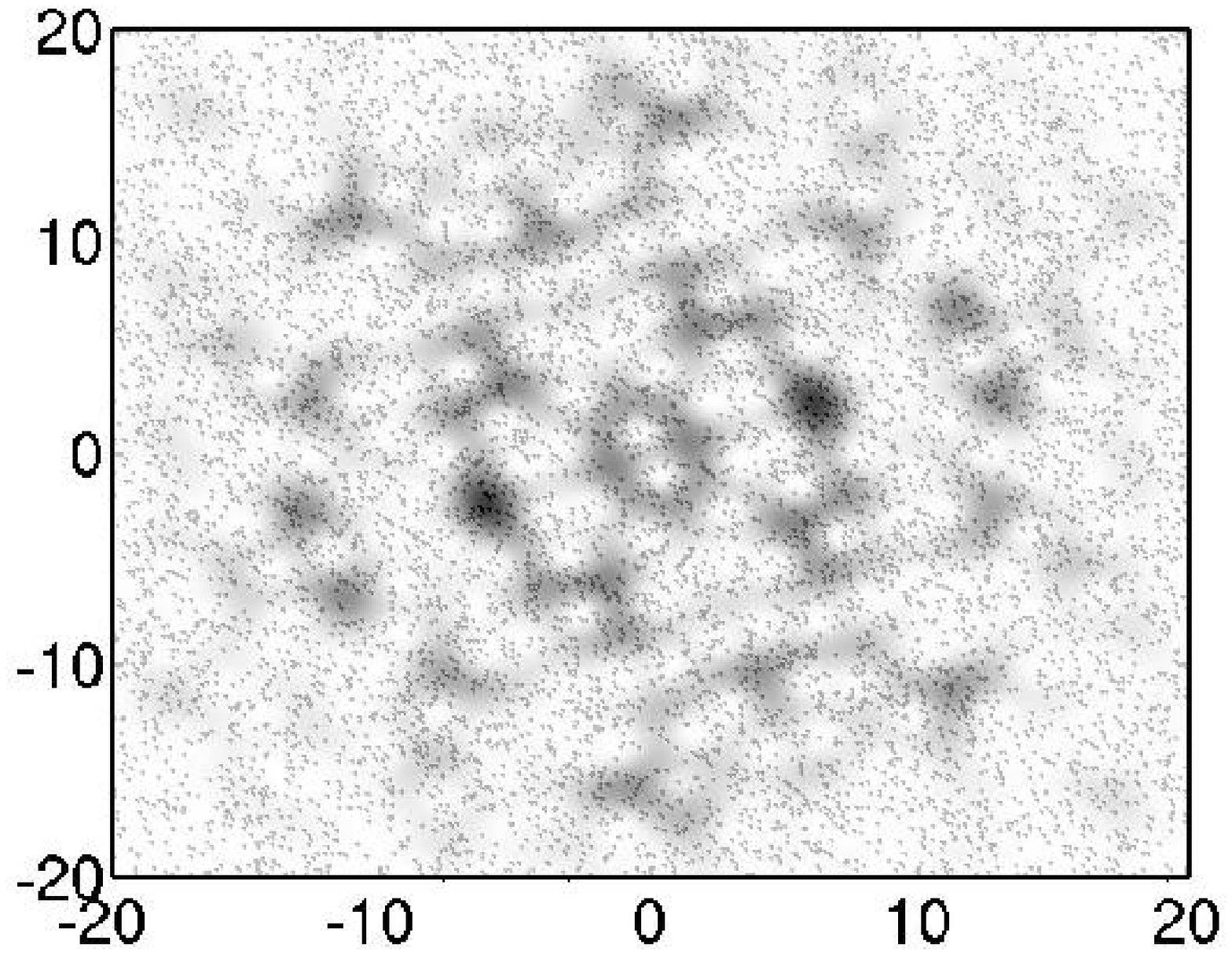}}
\put(6,5.5){\includegraphics[width=6cm]{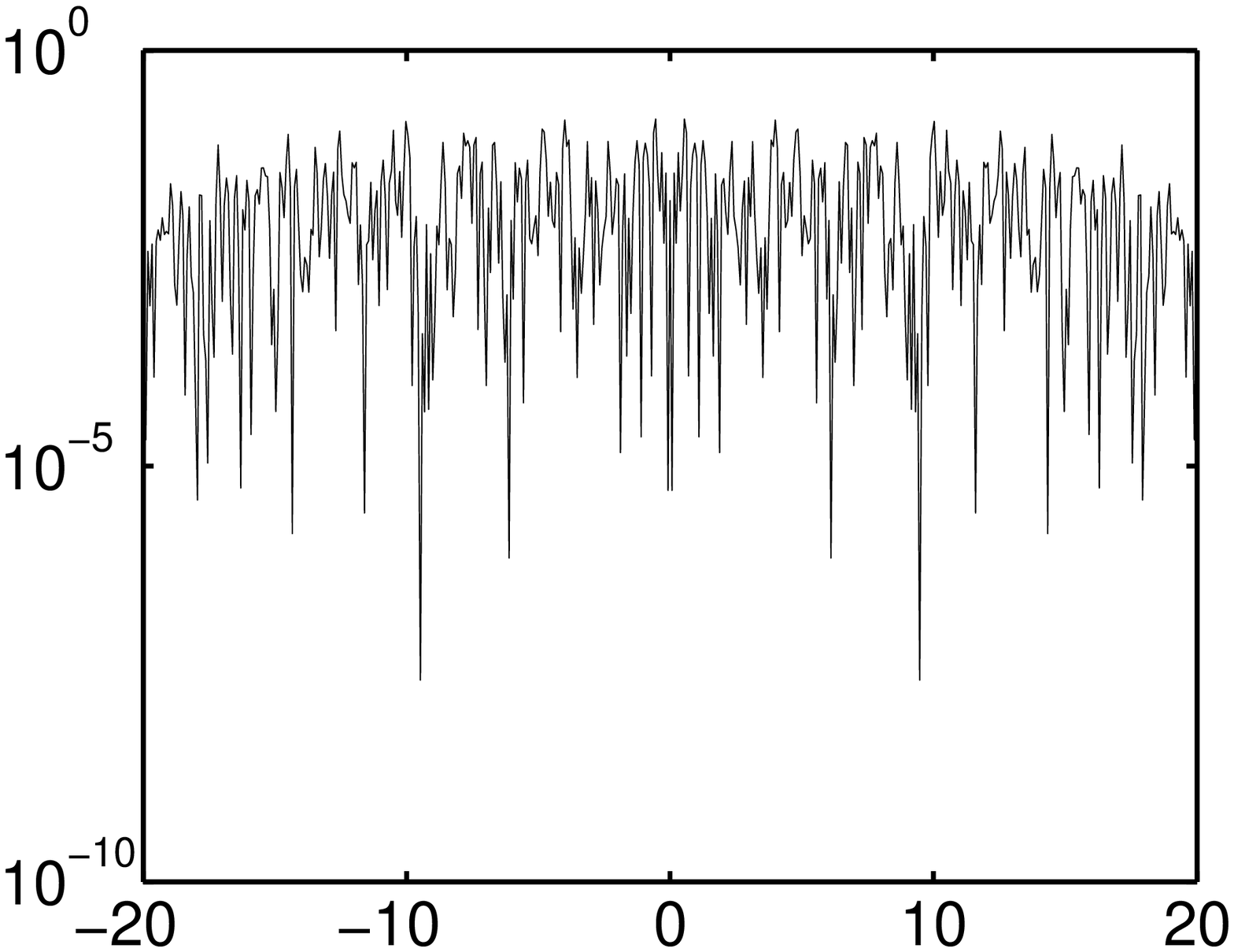}}
\put(6,0.5){\includegraphics[width=6.5cm]{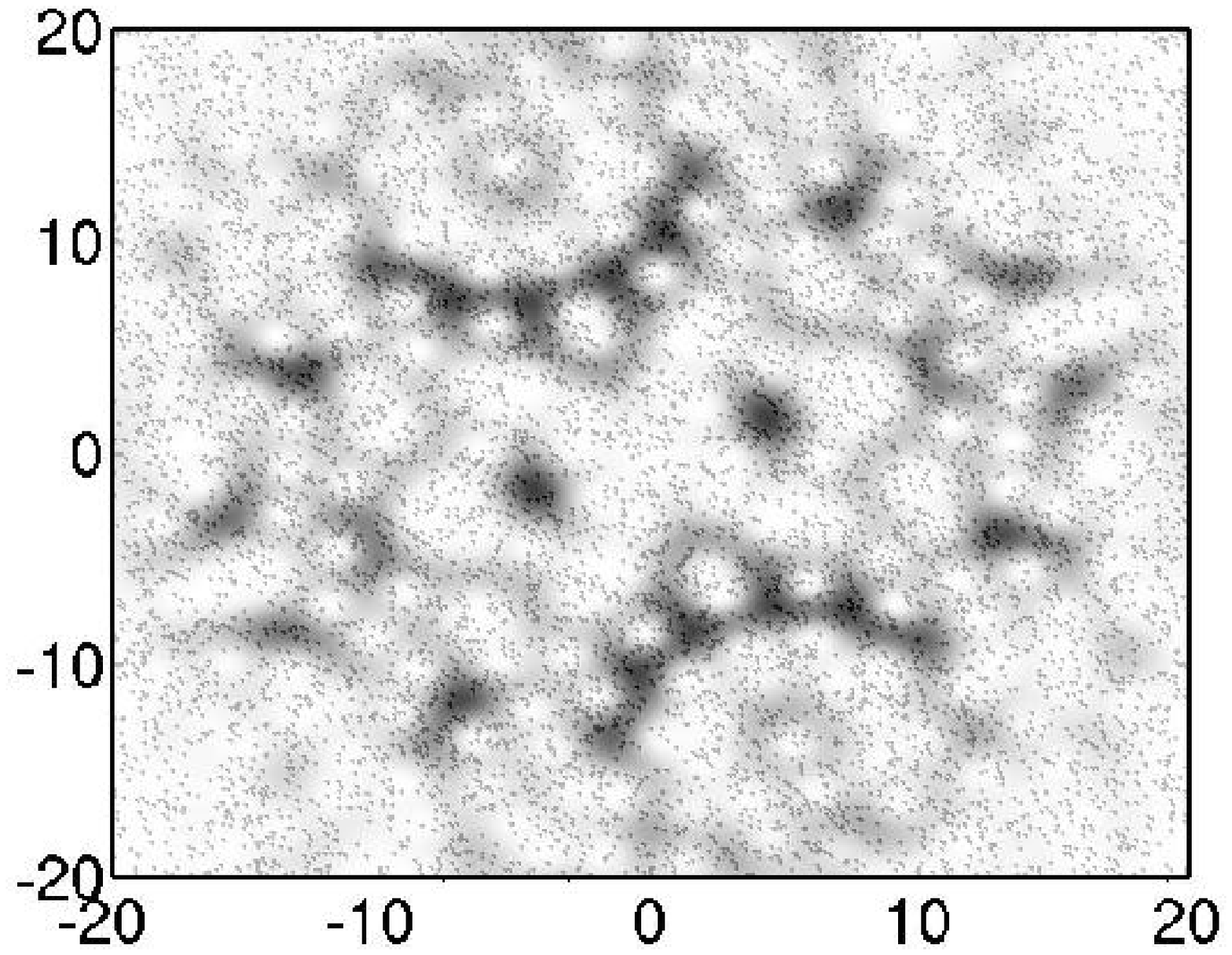}}
\put(0.5,9.5){\large(a)}
\put(7,9.5){\large(b)}
\put(0.5,4.5){\large(c)}
\put(7,4.5){\large(d)}
\put(-1.3,7.8){\large{$|\Psi|^2$}}
\put(-.7,3.0){\large{p}}
\put(2.75,0.4){\large{q}}
\put(9.25,0.4){\large{q}}
\end{picture}
\caption{\label{fig:states6}(a) and (b) Low energy eigenstates for the irrational system with kick strength $\mu = 6$, plotted on a logarithmic scale. Below each state, (c) and (d), are their Husimi functions plotted in overlay with the classical phase space Poincare plot. A Hilbert space of  dimension of $2^{10}$ was used for these simulations.}   
\end{center}
\end{figure}

As we have mentioned, using the definition of the Floquet operator given in Eq. (\ref{UF}), we can easily find a finite dimensional matrix that can then be diagonalised by brute force. This method has allowed us find eigenvalues and eigenvectors of matrices up to 3000x3000. To go larger requires more RAM. This can then be used to examine the structure of these eigenstates in detail as well as examining the system's eigenvalue statistics. As an example we show that, in the case of irrational kick frequencies, the Floquet operator contains some eigenstates that are clearly localised in space and some that appear to be extended. The existence of these states highlights what we have already found in the diffusion simulations. At low kick strength (say $\mu =1$), Fig. \ref{fig:states1}, an initial state may have eigenstate components which are localised. This is not the case however if we examine the eigenstates for the system with a larger kick strength. When $\mu =6$ even the lower energy eigenstates appear to have some sort of extended nature Fig. \ref{fig:states6}. These figures do not show all of the individual eigenstate. We have zoomed in on a small section to emphasis the Husimi function's structural similarity to classical phase map.

In conclusion, we have addressed the numerical problems associated with studying the quantum delta kicked harmonic oscillator. We have shown that the Fractional Fourier Transform may be used to infer crucial numerical information about the system. This includes improving the speed and accuracy with which we can evolve the system as well as giving an easy and effective way to analyse the system's Floquet eigenstates and quasi-energies. This improved numerical technique will assist in future studies of this systems dynamics using either a quantum information processor or the various physical systems mentioned in the introduction. 

\vspace{10mm}
\noindent
Acknowledgements: This research was supported by funding from Enterprise Ireland  under the research grant EI SC/2001/059.

\vspace{100mm}

\vspace{100mm}

\vspace{100mm}

\end{document}